\documentclass[10pt]{article}
\usepackage[cp1251]{inputenc}
\usepackage[english]{babel}

\title{{\bf \Large Modelling Tachyon Cosmology with Non-Minimal \\Derivative Coupling to Gravity}\\
{\normalsize ~~{\bf V.\,K. Shchigolev}\thanks{E-mail:
vkshch@yahoo.com},~~{\bf M.\,P. Rotova
\thanks{E-mail: canopus-007@yandex.ru} }}\\
{\small {\it Ulyanovsk State University, 42 L. Tolstoy Str.,
Ulyanovsk 432000, Russia}}\\
\bigskip
\small \begin{quote}{\bf Abstract} --  We study a tachyon model
with non-minimal derivative coupling to gravity in the
Friedmann-Robertson-Walker (FRW) flat cosmology. We propose the
special re-definition of the tachyon field which allows us to
represent tachyon field equation formally coinciding with its
usual form but with re-defined Hubble parameter. Two first
integrals for the model equations are obtained that can
essentially simplify both further analysis and analytical solving
for the model. These integrals become the trivial identities in
the case of minimal coupling. The effective energy density and
pressure of the tachyon field are obtained, and the necessary
condition of the possibility for this model to expand with
acceleration is derived. \\
\vspace{2,5mm}
{\bf PACS numbers:} 98.80.-k, 98.80.Es, 04.30.-w, 04.62.+v\\
{\bf Key words:} Cosmological Model, Non-Minimal Derivatives
Coupling, Tachyon Fields, Accelerated Expansion.\\
\end{quote}}
\date{}
\begin{document}

\maketitle \vspace{-2.5cm}

\section {Introduction}
\qquad According to recent observational data, such as SNe Ia
\cite{C1}-\cite{C3}, WMAP \cite{C4}, SDSS \cite{C5} etc., our
Universe is flat and undergoing a phase of the accelerated
expansion which started about five billion years ago. They also
suggest that it consists of about 70\% dark energy (DE) with
negative pressure, 30\% dust matter (cold dark matter plus
baryons), and negligible radiation. Dark energy has been one of
the most active fields in modern cosmology. In modern cosmology of
DE, the equation of state parameter $p=\gamma \rho$ plays an
important role, where $p$ and $\rho$ are its pressure and energy
density, respectively. To accelerate the expansion, the equation
of state must satisfy gamma $\gamma < -1/3$. As a possible
solution to the DE problem various dynamical models of DE have
been proposed, such as quintessence \cite{C6}, tachyon \cite{C7},
phantom \cite{C8} and quintom \cite{C9}, and so forth. Besides ,
other proposals on DE include interacting dark energy models
\cite{C10,C11}, braneworld models \cite{C12}, and holographic dark
energy models \cite{C13,C14}, etc. What distinguishes the tachyon
action from the standard Klein- Gordon form for scalar field is
that the tachyon action is non-standard and is of the
Dirac-Born-Infeld  form \cite{C7}.

In recent years, an alternative possibility of having an effective
scalar field theory governed by a Lagrangian containing a non
canonical kinetic term $ L = - V (\phi) F (X)$, where $X =
(1/2)\partial_{\mu}\phi \partial^{\mu}\phi$ has been proposed. One
of the most studied forms for $F(X)$ is $F(X) = \sqrt{1-2X}$. Such
a model can lead to late time acceleration of the Universe and is
called k-essence. This field can also give rise to inflation in
early universe and is called k-inflation. This type of field can
naturally arise in string theory and can be very interesting in
cosmological context \cite{C9}.

The alternative approach to the problem of accelerated expansion
is the consideration of various modifications  of the gravitation
theory. Among such modifications, the theories with non-minimal
coupling of a field to gravity are especially attractive. For
instance, scalar tensor theories are generalization of the
minimally coupled scalar field theories in a sense that here the
scalar field is non-minimally coupled with the gravity sector of
the action i.e with the Ricci scalar $R$. In these theories, the
scalar field participates in the gravitational interaction, unlike
its counterpart in the minimally coupled case where it behaves as
a non gravitational source like any other matter field.

Non-minimally coupled to gravity tachyon fields in cosmology have
been considered earlier, for example, in papers
\cite{C15}-\cite{C20}. In our paper we present the cosmological
paradigm with non-minimal {\it derivative} coupling between a
tachyon field and the curvature. Interest to such a type of
coupling between different fields and gravity is confirmed by
increasing number of publications devoted to this direction of
research (see, for example, \cite{C21}-\cite{C24}).  In addition
to the derivation of the main equations of the tachyon model with
non-minimal derivative coupling, we find the first integrals of
the model and conditions for accelerated expansion of the model.

\section{Main model equations}

\qquad  Let us consider a cosmological theory of a tachyon field
with non-minimal derivative coupling to curvature. In general, one
could have various forms of such coupling. For instance in the
case of two derivatives, one could have the terms $\xi R
\phi_{,\mu}\phi^{,\mu}$ and $\zeta R_{\mu \nu}
\phi^{,\mu}\phi^{,\nu}$, where the coefficients $\xi$, $\zeta$ are
the coupling parameters \cite{C21}.  Therefore, we start with the
following action:
\begin{equation}
\label {1} S=\int \left(-\frac{R}{2\kappa}-V(\phi)\sqrt
{\mathstrut 1-{\cal G} ^{\mu \nu}\phantom {.}\partial_{\mu} \phi
\phantom {.}\partial_{\nu} \phi} \phantom {.}\right)\sqrt{-g}~ d^4
x ,
\end{equation}
where $\kappa = 8 \pi G = M_P ^{-2}$ is the gravitational
constant, $M_P$ is the Planck mass, $ V (\phi)$ is a potential of
the tachyon field $ \phi $,
\begin{equation}
\label {2} {\cal G} ^{\mu \nu} =\alpha g^{\mu \nu}+\xi R g^{\mu
\nu} + \zeta R^{\mu \nu}.
\end{equation}
Here $\alpha, \xi$ and $\zeta$ are the dimensional constants of
the tachyon field coupling with gravity. For simplicity, we
consider a spatially flat FRW cosmological model with the
space-time interval
\begin{equation}
\label {3} d s^2 = N^2(t) d t^2- a^2 (t)\left(d r^2+r^2d \Omega
^2\right),
\end{equation}
where $a(t)$ is a scale factor, and $N(t)$ is a lapse function.
Assuming homogeneity of the tachyon field, i.e. $\phi =\phi(t)$,
and calculating for the interval (\ref{3})
\begin{equation}
\label {4} R_{00}= - \frac{3 \ddot a N - \dot a \dot N}{aN},~~R =
-6 \frac{a\ddot a N-a \dot a \dot N+\dot a^2 N}{a^2N^3},
\end{equation}
from (\ref{1}) one can obtain  the following expression for the
kinetic term $X ={\cal G}^{\ mu \
nu}\,\partial_{\mu}\phi\,\partial_{\nu}\phi$:
\begin{equation}
\label {5} X = \dot \phi ^2
\left[\frac{\alpha}{N^2}-3(2\xi+\zeta)\frac{\ddot a}{a N^4}+(6 \xi
+ \zeta)\frac{\dot a \dot N}{a N^5}-6 \xi \frac{\dot a^2}{a^2 N^4}
\right].
\end{equation}

On account of the latter and metrics (\ref {3}) we obtain from
action (\ref {1}) the effective Lagrangian of the model in the
form:
\begin{equation}
\label {6} L = \frac{3}{8\pi G} \left(\frac{a^2\ddot
a}{N}+\frac{a\dot a^2}{N}-\frac{a^2\dot a \dot N}{N^2}\right)- a^3
N V(\phi)\sqrt{\mathstrut 1-X}{~~}.
\end{equation}

As one can see, the term proportional to the second derivative
$\ddot a $ is in $X$ of (\ref {5}), i.e. in the term non-linearly
included in Lagrangian (\ref {6}). It means that the dynamical
equations of the model will contain the scale factor derivatives
of the order higher than two. As it was noted in \cite{C21} for
the canonical non-minimal scalar field theory, the appearance of
such derivatives can be avoided due to the particular choice of
coupling constants in (\ref {2}) without loss of generality.
Therefore, we also put $\displaystyle \xi = - \frac {1}{2} \zeta$.
The latter means that tensor (\ref {2}) is expressed through the
metrics tensor and the Einstein tensor as
$$
{\cal G} ^{\mu \nu} =\alpha g^{\mu \nu}+ \zeta\Big( R^{\mu
\nu}-\frac{1}{2}g^{\mu \nu}R\Big).
$$
The corresponding expression for $X$ in (\ref {5}) can be
re-written as follows:
\begin{equation}
\label {7} X = \frac{\dot \phi ^2}{N^2}Y,~~Y = \alpha- 2
\zeta\frac{\dot a \dot N}{a N^3}+3 \zeta \frac{\dot a^2}{a^2 N^2}.
\end{equation}

Integrating the first term in Lagrangian (\ref {6}) by parts and
substituting (\ ref{7}),  we obtain the following effective
Lagrangian of the model:
\begin{equation}
\label {8} L = -\frac{3}{8\pi G} \frac{a\dot a^2}{N}- a^3
V(\phi)\sqrt{\mathstrut N^2-\dot \phi^2 \, Y}.
\end{equation}

Then taking into account the explicit form of $Y$ from (\ref {7})
and putting the gauge $N = 1$, we obtain by variation over $ \phi,
N$ and $a$ the following main equations of the model:
\begin{equation}
\label{9} \frac{\ddot \phi Z}{1-\dot \phi^2 Z}+3H\dot\phi \Bigl(
Z+\zeta \dot H \,\frac{2-\dot \phi ^2 Z}{1-\dot \phi ^2 Z}\Bigr) +
\frac{V'}{V}=0 ,
\end{equation}
\begin{equation}
\label{10} \frac{3}{8\pi G}\,H^2 =  V\frac{1-\zeta(\dot\phi^2 \dot
H +2 \dot\phi \ddot\phi H)}{\sqrt{1-\dot \phi^2 Z}}-V'\zeta\frac{
\dot\phi^3 H}{\sqrt{1-\dot \phi^2 Z}}-V\zeta \dot \phi^2 H
\frac{\displaystyle\dot\phi \ddot\phi Z + 3 \zeta \dot\phi^2 H
\dot H }{(1-\dot \phi^2 Z)^{3/2}} ,
\end{equation}
\begin{eqnarray}
\frac{1}{8\pi G}(2\dot H + 3 H^2)= V \sqrt{1-\dot \phi^2 Z} +V
\zeta \frac{3\dot \phi^2
H^2 + (\dot \phi^2 \dot H + 2 \dot\phi \ddot\phi H)}{\sqrt{1-\dot \phi^2 Z}}+  \nonumber\\
+V'\zeta\frac{ \dot\phi^3 H}{\sqrt{1-\dot \phi^2 Z}}+ V\zeta \dot
\phi^2 H \frac{\displaystyle\dot\phi \ddot\phi Z + 3 \zeta
\dot\phi^2 H \dot H }{(1-\dot \phi^2 Z)^{3/2}} ,\label{11}
\end{eqnarray}
where $Z = Y_{\mid N=1}$, that is
\begin{equation}
\label{12} Z= \alpha+ 3\zeta H^2,
\end{equation}
where $H = \dot a/a$ is the Hubble parameter, $V'=dV/d\phi$.

By addition of equations (\ref {10}) and (\ref {11}), one can
obtain:

\begin{equation}
\label{13}\frac{1}{4\pi G}(\dot H + 3 H^2)= V \sqrt{1-\dot \phi^2
Z} + V\,\frac{1+3\zeta H^2\dot \phi ^2 }{\sqrt{1-\dot \phi^2 Z}}.
\end{equation}

Transforming the tachyon field $\phi(t) \to \Phi (t)$ according to
\begin{equation}
\label{14} \dot \phi \sqrt{Z} = \dot \Phi,
\end{equation}
and using the consequent expression
\begin{equation}
\label{15} \dot\phi \ddot\phi Z + 3 \zeta \dot\phi^2 H \dot H =
\dot \Phi \ddot \Phi,~~ \frac{d V}{d \phi} = \sqrt{Z}\,\frac{d
V}{d \Phi},
\end{equation}
we can re-write (\ref {9}) for the tachyon field $\Phi$ as
follows:
\begin{equation}
\label{16} \frac{\ddot \Phi}{1-\dot \Phi^2}+3H\dot\Phi \Bigl(
1+\frac{\zeta \dot H}{\alpha+ 3\zeta H^2}\Bigr) + \frac{V'}{V}=0,
\end{equation}
where $V' = \displaystyle \frac{dV}{d\Phi}$.

Equation (\ref{16}) differs from the tachyon equation in the case
of minimal coupling only in the presence of the second term in
brackets, and reduces to it as $\zeta = 0$ or $\dot H = 0$. The
latter means either the minimally coupled tachyon field or the
constancy of the Hubble parameter.  In view of $Y$ from (\ref {7})
for $H=constant$, only re-scaling of parameter $\alpha$ in
Lagrangian (\ref {8}) is required to remain in the minimally
coupled theory. Complete similarity of equation (\ref {16}) with
the similar equation in the case of minimal coupling and $H\neq
constant$ can be obtained by substitution
\begin{equation}\label{17}
f = a\, Z^{1/6},~~~{\cal H} = \frac{\dot f}{f} = H+\frac{\dot Z}{6
Z},
\end{equation}
which transforms equation (\ref{16}) as follows:
\begin{equation}\label{18}
\frac{\ddot \Phi}{1-\dot \Phi^2}+3{\cal H} \dot\Phi  +
\frac{V'}{V}=0.
\end{equation}
Substituting  (\ref {14}), (\ref {15}) into equations (\ref {10})
and (\ref {13})  we can obtain the model equations as follows:
\begin{equation}\label{19}
\frac{3}{8\pi G}\,H^2 =  \frac{V}{\sqrt{1-\dot \Phi^2
}}+V\zeta\frac{3{\cal H}\dot \Phi^2\Bigl (\displaystyle \dot
\Phi^2 H Z^{-1}\Bigr)- \frac{d}{dt}\Bigl ( \dot \Phi^2 H
Z^{-1}\Bigr)}{\sqrt{1-\dot \Phi^2}},
\end{equation}
\begin{equation}\label{20}
\frac{1}{4\pi G}(\dot H + 3 H^2)= V \sqrt{1-\dot \Phi^2} +
V\,\frac{1+3\zeta H^2 Z^{-1}\dot \Phi ^2}{\sqrt{1-\dot \Phi^2}},
\end{equation}
The set of equations (\ref {18}) - (\ref {20}) is rather
complicated to be analyzed and exactly solved. However, the
situation is essentially simplified if we take into account the
existence of the first integrals. In the next section, we get two
such integrals. We also briefly discuss the effective equation of
state for the model.

\section{The first integrals and equation of state}

\qquad First of all, we will show that equation (\ref {11}) is a
differential consequence of the following first integral:
\begin{equation}
\label{21} \frac{3}{8\pi G}H^2 - V\,\frac{1+3\zeta H^2\dot \phi ^2
}{\sqrt{1-\dot \phi^2 Z}}=\frac{K}{a^3},
\end{equation}
where $K = constant$, and the tachyon field equation (\ref{9}).
Indeed, multiplying equation (\ref {21}) by $a ^ 3$ and
differentiating the result with respect to time, we have:
$$
\frac{3}{8\pi G}(\dot a^3+ 2 a \dot a \ddot a)= V' \dot \phi a^3
\frac{1+3\zeta H^2 \dot \phi^2}{\sqrt{1-\dot \phi^2 Z}}+3Va^2\dot
a \frac{1+3\zeta H^2\dot \phi ^2}{\sqrt{1-\dot \phi^2 Z}}+
$$
$$
+3Va^3 \zeta \frac{2H\dot H \dot \phi^2 + 2 H^2 \dot \phi \ddot
\phi}{\sqrt{1-\dot \phi^2 Z}}+Va^3(1+3\zeta H^2\dot \phi
^2)\frac{\dot \phi \ddot \phi Z+ \dot \phi^2 3 \zeta H \dot
H}{(1-\dot \phi^2 Z)^{3/2}}=0.
$$
Dividing the latter by $a ^ 3 $ and using the identity $ \ddot a /
a = \dot H + H ^ 2 $, we get:
$$
\frac{3 H}{8 \pi G}(2\dot H + 3 H^2)= 3 H \Bigl\{ V \sqrt{1-\dot
\phi^2 Z} +V \zeta \frac{3\dot \phi^2 H^2 + (\dot \phi^2 \dot H +
2 \dot\phi \ddot\phi H)}{\sqrt{1-\dot \phi^2 Z}}+ V'\zeta\frac{
\dot\phi^3 H}{\sqrt{1-\dot \phi^2 Z}}+
$$
$$
+V\zeta \dot \phi^2 H \frac{\displaystyle\dot\phi \ddot\phi Z + 3
\zeta \dot\phi^2 H \dot H }{(1-\dot \phi^2
Z)^{3/2}}\Bigr\}+\Bigl[V\frac{\dot\phi \ddot\phi Z + 3 \zeta
\dot\phi^2 H \dot H }{(1-\dot \phi^2 Z)^{3/2}}+V\frac{3\zeta H
\dot H \dot \phi ^2}{\sqrt{1-\dot \phi^2 Z}}+\frac{3 V
H}{\sqrt{1-\dot \phi^2 Z}}+
$$
$$
+\frac{V' \dot \phi}{\sqrt{1-\dot \phi^2 Z}}-3V H \sqrt{1-\dot
\phi^2 Z}\Bigr].
$$
The expression in the square brackets is zero in view of equation
(\ref {9}). On cancelling out $3 H$ in the remaining equation, one
is left with equation (\ref {11}). In terms of re-defined field
\ref {14}), the first integral (\ref {21}) can be written as:
\begin{equation}
\label{22} \frac{3}{8\pi G}H^2 = V\,\frac{1+3\zeta H^2 Z^{-1}\dot
\Phi ^2}{\sqrt{1-\dot \Phi^2}}+\frac{K}{a^3},
\end{equation}
In the case of minimal coupling with $ \alpha = 1, ~ \zeta = 0, ~
Z = 1 $, equation (\ref {21}) coincides with equation (\ref {10})
considered under the same conditions and zero constant of
integration $ K = 0 $. The latter follows from the existence of
only two independent equations in minimally coupled model.

From equations (\ref {20}) and (\ref {22}), the following
consequence can be obtained:
\begin{equation}\label{23}
\frac{3}{4\pi G} \frac{\ddot a}{a} \equiv \frac{3}{4\pi G}(\dot H
+ H^2)= V\,\frac{\displaystyle 2-3(1+\zeta H^2 Z^{-1})\dot \Phi ^2
}{\sqrt{1-\dot \Phi^2}}-4\, \frac{K}{a^3}.
\end{equation}
This equation is convenient to analyze the possibility of
accelerated expansion  $\ddot a>0$.

On the other hand, comparing the right sides of (\ref {20}) and
(\ref {22}) with similar equations for a perfect fluid (i.e. $ -
(\rho + 3 p)$ and $ \rho $, respectively ), we can obtain the
following formal expressions for the effective energy density and
pressure:
\begin{equation}\label{24}
\rho = V\, \frac{1+3\zeta H^2 Z^{-1}\dot \Phi ^2}{\sqrt{1-\dot
\Phi^2 }}+\frac{K}{a^3},~~~~p = -V\,\sqrt{1-\dot
\Phi^2}+\frac{4}{3}\,\frac{K}{a^3}.
\end{equation}

These expressions imply the effective equation of state of the
form:
\begin{equation}\label{25}
\gamma \equiv \frac{p}{\rho} = -\,\frac{3a^3 V(1-\dot
\Phi^2)-4K\sqrt{1-\dot \Phi^2}}{3a^3 V(1+3\zeta H^2 Z^{-1}\dot
\Phi ^2)+3K\sqrt{1-\dot \Phi^2}},
\end{equation}
which can be written as
\begin{equation}\label{26}
\gamma _{K=0}= -\,\frac{1-\dot \Phi^2}{1+3\zeta H^2 Z^{-1}\dot
\Phi ^2}
\end{equation}
in the case of zero constant $ K = 0 $.

From the condition of accelerated expansion,  $ \gamma _ {K = 0}
<-1 / 3 $, we have the following inequality
\begin{equation}\label{27}
3(1+\zeta H^2 Z^{-1})\dot \Phi ^2 < 2,
\end{equation}
This condition can also be obtained from $ \ddot a> 0 $ in
equation (\ref {23}). It should be noted that due to uncertainty
in the sign of constant $ K \neq 0 $ in (\ref {25}) the additional
terms $ \sim K $ can significantly change equation of state in
each direction from the critical value $ \gamma =- 1 $.

Another first integral follows from the equality of the
right-hand-sides of equations (\ref{19}) and (\ref {22}). After
the substitution of $ 3 {\cal H} \dot \Phi $ from equation (\ref
{18}), the result can be written as
$$
\zeta \, \frac{d}{dt}\ln\Bigl[ \frac{\dot \Phi^2 H Z^{-1}a^3
V}{\sqrt{1-\dot \Phi ^2}}\Bigr]+\frac{K\sqrt{1-\dot \Phi
^2}}{a^3\,\dot \Phi^2 H Z^{-1} V}=0.
$$
Hence, we obtain the following first integral:
\begin{equation}\label{28}
\zeta\,\Bigl(\frac{\dot \Phi^2 H a^3
}{Z}\Bigr)\,\frac{V}{\sqrt{1-\dot \Phi ^2}}=-K \,t+\zeta\, L,
\end{equation}
where $ L $ is an arbitrary constant. In the case of minimal
coupling, i.e. for $ \zeta = 0 $, we obtain $ K = 0 $, that is a
trivial identity. The latter means the absence of the first
integral of the form (\ref {28}) in the case of minimal coupling
to gravity. From (\ref {28}), it also follows that $ K = L = 0 $
for $ V = 0 $ and / or $ \dot \Phi = 0 $. In the first case, we
obtain from (\ref {24}) the expected result : $ \rho = p = 0 $. In
the second case from (\ref {26}),  we obtain the quasi vacuum
state $ \gamma = -1 $ with $ p = - \rho = - V_0 = constant $,
because of equation (\ref {18}) implies the constancy of potential
$ V = V_0 $ for $ \dot \Phi = 0 $.

Possessing these first integrals, we can try to construct some
exact solutions of the model. One can offer a possible way for
that which consists of the following. For simplicity, we assume
that $ K = 0 $ and substitute expression  $ V / \sqrt {1 - \dot
\Phi ^ 2} $ from (\ref {28}) into equation (\ref {22}). As a
result, we have:
\begin{equation}\label{29}
\frac{3}{8\pi G}H^2 = \frac{L Z}{a^3 \dot \Phi^2
H}\,\Bigl(1+3\zeta H\,\frac {\dot \Phi ^2 H}{Z}\Bigr).
\end{equation}
By specifying a certain dependence $ \dot \Phi(t) $, one can find
$ H(t) $ and $ a(t) $ from this equation. Let us note that due to
(\ref {24}) equation (\ref {29})  can be expressed in terms of the
original field $ \phi $ in (\ref {14}) as
$$
\frac{3}{8\pi G}\dot a^3 = L\,\Bigl(\frac{1}{\dot \phi^2}+3\zeta
\,\frac {\dot a}{a}\Bigr).
$$

After that, it becomes possible to obtain a potential $ V $ from
(\ref {28}) or (\ref {18}). Obviously, we have to be confident
that these two results for $V$ will coincide. So differentiating
(\ref {28}) and partially replacing the terms with the help of
equation (\ref {18}), after simple manipulations one can show that
these two solutions for $V(\Phi)$ will coincide if the following
equality is valid:
\begin{equation}\label{30}
\frac{d}{d t}\ln \Bigl(\frac{\dot \Phi^2 H a^3 }{Z}\Bigr) = 3
{\cal H } \dot \Phi^2.
\end{equation}

Substituting $\displaystyle \frac{\dot \Phi^2 H a^3 }{Z}$ from
(\ref {28}) with $K=0$ into (\ref {30}) we have
\begin{equation}\label{31}
\frac{d}{dt}\Bigl(\ln \frac{V}{\sqrt{1-\dot \Phi^2}}\Bigr)+ 3 {\cal H}
\dot \Phi^2 =0.
\end{equation}
Obviously, the latter is just the tachyon field equation (\ref
{18}).

Of course, there exist different approaches for finding the exact
solutions but (\ref {30}) always  has to be taken  into account in
the case of non-minimal coupling. We have to emphasize that there
is no need to verify this equation in the case of minimal coupling
because the first integral (\ref {28}) does not exist at all as
$\zeta = 0$. Nevertheless, the minimally coupled analog of
equation (\ref {31}) is valid, that can be seen from the usual
tachyon equation:
\begin{equation}\label{32}
\frac{\ddot \phi}{1-\dot \phi^2}+3 H \dot\phi  + \frac{V'}{V}=0
\,\,\,\Leftrightarrow \,\,\,
\frac{d}{dt}\Bigl(\ln \frac{V}{\sqrt{1-\dot \phi^2}}\Bigr)+ 3 H \dot
\phi^2 =0.
\end{equation}
It is appropriate mention here that, on view of the Friedmann
equation
\begin{equation}\label{33}
\frac{3}{8\pi G}\,H^2 =  \frac{V}{\sqrt{1-\dot \phi^2 Z}},
\end{equation}
and (\ref {32}),  we can find the following formulas:
$$
\dot \phi ^2(t) =
\frac{2}{3}\frac{d}{dt}\Bigl(\frac{1}{H}\Bigr),~~~~~V(t)=
\frac{3}{8\pi
G}\,H^2\,\sqrt{1-\frac{2}{3}\frac{d}{dt}\Bigl(\frac{1}{H}\Bigr)},
$$
which allows us, starting with some $H(t)$ given, to get a wide
class of exact solutions for the case of minimal coupling. The
alternative approach serving this same purpose is recently
proposed in \cite{C25}. Unfortunately, it is not so easy to obtain
the exact solutions in the case of non-minimal derivative
coupling.

\section{Conclusion}

\qquad Thus, we have studied  the spatially  flat FRW tachyon
model with non-minimal derivative coupling to gravity. We have
obtained the main equations of the model in the form (\ref {9}) -
(\ref {11}). Besides, it was suggested that the specific
representation of tachyon field (\ref {14}) could allow us to get
the equation formally identical to a similar equation with minimal
coupling (\ref {18}), but with the re-defined Hubble parameter
(\ref {17}). Two first integrals (\ref {22}), (\ref {28}) for the
model equations have been found which could essentially simplify
both analysis and solution,  and which became the identities in
the case of minimal coupling. The set of main equations governing
the model consists of (\ref {22}), (\ref {28}) and (\ref {30}) at
least in the case of $K=0$. Thus, we have proposed a method for
constructing exact solutions for the tachyon model in FRW
cosmology based on the first integrals obtained above. In the
framework of our model, the effective energy density and pressure
of the tachyon field (\ref {24}) have been found. The conditions
for the cosmic accelerated expansion (\ref {27}) have been found
as well. Further details and consequences of the model considered
here are in progress.

\end{document}